\documentstyle{article}
\def\be{\begin{equation}}
\def\ee{\end{equation}}
\def\bea{\begin{eqnarray}}
\def\eea{\end{eqnarray}}
\def\nn{\nonumber}
\def\ba{\begin{array}}
\def\ea{\end{array}}   
\def\zero{0 \hskip -2mm 0}
\def\one{1\hskip -1mm{\rm l}}
\def\id{1\hskip -0.8mm{\rm l}} 
\def\half{\frac{1}{2}}  
\def\cE{{\cal E}}
\def\cH{{\cal H}}
\def\cO{{\cal O}}
\def\hO{\hat{O}}
\def\hOA{\hat{O}_A}
\def\cHA{\cH_A}
\def\al{\alpha}
\def\eps{\epsilon}
\def\g{\gamma}
\def\psia{\psi_A}
\def\A{{\mbox{\boldmath $A$}}}
\def\B{{\mbox{\boldmath $B$}}} 
\def\S{{\mbox{\boldmath $S$}}}
\def\p{{\mbox{\boldmath $p$}}}
\def\r{{\mbox{\boldmath $r$}}}
\def\zp{z^\prime}
\def\val{{\mbox{\boldmath $\alpha$}}}
\def\vnab{{\mbox{\boldmath $\nabla$}}}
\def\Vomeg{\underline{\mbox {\boldmath $\Omega$}}}
\def\Omeg{{\mbox {\boldmath $\Omega$}}} 
\def\vpi{\mbox{\boldmath $\pi$}}
\def\vsig{{\mbox{\boldmath $\sigma$}}} 
\def\Vsig{{\mbox {\boldmath $\Sigma$ }}} 

\def\hpi{\hat{\pi}}
\def\hvpi{\hat{\vpi}}
\def\hvpip{\hvpi_\perp}
\def\curl{{\rm curl}}
\def\ih{i\hbar}
\def\ddz{\frac{\partial}{\partial z}} 
\def\rp{\r_\perp}
\def\pp{\p_\perp}
\def\hp{\hat{\p}}
\def\hpp{\hp_\perp}  
\def\hppsq{\hat{p}_\perp^{\,2}}

\def\uzpz{\hat{U}(\zp,z)}
\def\uzpzd{\uzpz^\dagger}
\def\xin{\langle x \rangle_i}
\def\yin{\langle y \rangle_i}
\def\xf{\langle x \rangle_f}
\def\yf{\langle y \rangle_f}
\def\pxi{\left\langle \hat{p}_x \right\rangle_i}
\def\pyi{\left\langle \hat{p}_y \right\rangle_i}
\def\pxf{\left\langle \hat{p}_x \right\rangle_f}
\def\pyf{\left\langle \hat{p}_y \right\rangle_f}
\def\sxi{\left\langle S_x \right\rangle_i}
\def\syi{\left\langle S_y \right\rangle_i}
\def\szi{\left\langle S_z \right\rangle_i}
\def\sxf{\left\langle S_x \right\rangle_f}
\def\syf{\left\langle S_y \right\rangle_f}
\def\szf{\left\langle S_z \right\rangle_f}
\def\sixi{\left\langle \sigma_x \right\rangle_i}
\def\siyi{\left\langle \sigma_y \right\rangle_i}
\def\xsxi{\left\langle x S_x \right\rangle_i}

\def\ysyi{\left\langle y S_y \right\rangle_i}
\def\xszi{\left\langle x S_z \right\rangle_i}
\def\yszi{\left\langle y S_z \right\rangle_i}
\def\pxsxi{\left\langle \hat{p}_x S_x \right\rangle_i}

\def\pysyi{\left\langle \hat{p}_y S_y \right\rangle_i}
\def\pxszi{\left\langle \hat{p}_x S_z \right\rangle_i}
\def\pyszi{\left\langle \hat{p}_y S_z \right\rangle_i}
\begin{document}
\begin{flushright}
physics/9803042 
\end{flushright}
\begin{flushright}
IMSc/98/03/12
\footnote{Presented at the Workshop on 
``Quantum Aspects of Beam Physics'', 15th ICFA (International 
Committee for Future Accelerators) Advanced Beam Dynamics Workshop, 
January 4 - 9, 1998, Monterey, California, U.S.A.  To appear in the 
Proceedings of the Workshop, Ed. Pisin Chen (World Scientific, 
Singapore, 1998).}
\end{flushright}

\begin{center}

{\large \bf THE DIRAC EQUATION APPROACH TO \\ 
SPIN-$\half$ PARTICLE BEAM OPTICS} 

\medskip

{R. JAGANNATHAN} 

{\em The Institute of Mathematical Sciences \\
4th Cross Road, Central Institutes of Technology Campus \\ 
Tharamani, Chennai $($Madras$)$, Tamilnadu - 600 113, INDIA} \\
E-mail: {\tt jagan@imsc.ernet.in} \\ 
{\tt http://www.imsc.ernet.in/}$\sim${\tt jagan} 

\end{center}

\bigskip

\begin{quote}
{The traditional approach to accelerator optics, based mainly on 
classical mechanics, is working excellently from the practical point of 
view.  However, from the point of view of curiosity, as well as with a 
view to explore quantitatively the consequences of possible small 
quantum corrections to the classical theory, a quantum mechanical 
formalism of accelerator optics for the Dirac particle is being 
developed recently.  Here, the essential features of such a quantum 
beam optical formalism for any spin-$\half$ particle are reviewed.  
It seems that the quantum corrections, particularly those due to the 
Heisenberg uncertainty, could be important in understanding the 
nonlinear dynamics of accelerator beams.}
\end{quote}

\newpage

\section{Introduction}
Why should one bother about a quantum mechanical treatment of  
accelerator {\em optics} when the classical treatment works so well?  
This is a natural question indeed.  There is no {\em prima facie} 
reason at all to believe that a quantum mechanical treatment would be 
necessary to deal with any aspect of accelerator {\em optics} design.  
As has been rightly pointed out~\cite{C}, primary effects in conventional 
accelerators are essentially classical since the de Broglie wavelength of 
the high energy beam particle is much too small compared to the typical 
apertures and the energy radiated is typically low and of long  
wavelength.  However, it is being slowly recognized that quantum effects 
are still important due to demands on high precision in accelerator 
performance and ever increasing demands for higher beam energy,  
luminosity and brightness~\cite{C}.  Also, there is a growing feeling 
now that a complete picture of spin polarization can only be obtained on 
the basis of coupled spin $\leftrightarrow$ orbit phase-space transport 
equations (with and without radiation) and to include all the subtleties 
of radiation one has to begin with quantum mechanics since classical 
white noise models simply would not suffice for all situations~\cite{HB}.  
I like to add: 

\begin{itemize}

\item After all, accelerator beam is a quantum mechanical system 
and one may be curious to know how its classical behavior can be 
understood, in detail, from a quantum mechanical formalism based 
on the appropriate relativistic wave equation.  

\item As has been revealed by recent studies~\cite{Ca} the passage from 
quantum theory to classical theory is not a simple affair, particularly 
when the system is a complicated nonlinear dynamical system with regular 
and chaotic regions in its phase-space.  Since accelerator beams are such 
systems~\cite{M} it is time that the quantum mechanics of accelerator 
optics is looked at seriously.  

\end{itemize}

Essentially from the point of view of curiosity, the axially symmetric 
magnetic lens was first studied~\cite{JSSM} based completely on the 
Dirac equation.  Later works~\cite{J,JK1} led to further insights 
into the quantum mechanics of spin-$\half$ particle beam optics.  
Quantum mechanics of the Klein-Gordon (spin-$0$) and 
nonrelativistic Schr\"{o}dinger charged-particle beams were also 
studied~\cite{JK1,KJ}.  These works dealt essentially with aspects 
of ion optics and electron optical imaging (for an excellent survey 
of scalar electron wave optics see the third volume of the 
encyclopaedic three-volume text book of Hawkes \& Kasper~\cite{HK}; 
this contains also references to earlier works on the use of the 
Dirac equation in electron wave optics problems like diffraction, 
to take into account the spinor nature of the electron).  

In the context of accelerator physics also, like in the case of 
electron and ion beam optical device technologies, the practice of 
design of beam optical elements is based mainly on classical physics.  
As is well known, various aspects of accelerator beam dynamics 
like orbital motion, spin evolution and beam polarization, radiation 
and quantum fluctuations of trajectories, are analyzed piecewise using 
classical, semiclassical, or quantum theories, or a mixture of them, 
depending on the situation treated.  Quantum mechanical implications 
for low energy polarized (anti)proton beams in a spin-splitter device, 
using the transverse Stern-Gerlach~(SG) kicks, have been  
analyzed~\cite{CP} on the basis of nonrelativistic Schr\"{o}dinger 
equation.    

To obtain the coupled spin $\leftrightarrow$ phase-space transport 
equations, for the spin-$\half$ particle, one needs an appropriate 
quantum Hamiltonian.  Such a Hamiltonian was stated, as following from 
the systematic Foldy-Wouthuysen~(FW) transformation technique~\cite{FW}, 
by Derbenev and Kondratenko~\cite{DK} in 1973 as the starting point of 
their radiation calculations but no explicit construction was 
given (such a Hamiltonian can also be justified~\cite{Ja} using the 
Pauli reduction of the Dirac theory).  The Derbenev-Kondratenko~(DK) 
Hamiltonian has been used~\cite{BHR} to construct a completely 
classical approach to beam optics, including spin components as 
classical variables.  Now, a detailed derivation of the DK Hamiltonian 
has been given~\cite{HB} and a completely quantum mechanical formalism 
is being developed~\cite{HB} in terms of the `machine coordinates' and 
`observables', suitable for treating the physics of spin-$\half$ 
polarized beams from the point of view of machine design. 

Independent of the DK formalism, recently~\cite{CJKP1,JK2} we have 
made a beginning in the application of the formalism of the Dirac spinor 
beam optics, developed earlier~(\cite{JSSM}-\cite{JK1}) mostly with 
reference to electron microscopy, to accelerator optics to understand in 
a unified way the orbital motion, SG kicks, and the 
Thomas-Bargmann-Michel-Telegdi~(TBMT) spin evolution.  Here, I present 
the essential features of our work, done so far, on the quantum beam 
optical approach to accelerator optics of spin-$\half$ particles based 
on the Dirac equation.  

\section{Quantum beam optics of the Dirac particle}
Our formalism of quantum beam optics of the Dirac particle, in the 
context of accelerators, is only in the beginning stages of 
development.  So, naturally there are several simplifying 
assumptions: We deal only with the single particle dynamics, based 
on the single particle interpretation of the Dirac equation, 
ignoring all the inter-particle interactions and statistical 
aspects.  Only monoenergetic beam is considered.  The treatment is 
at the level of paraxial approximation, so far, though the general 
framework of the theory is such that extension to the case of 
nonparaxial (nonlinear) systems is straightforward.  Only 
time-independent magnetic optical elements with straight axis are 
considered.  Electromagnetic field is treated as classical.  And, 
radiation is ignored.  

Thus, we are dealing with elastic scattering of the particles of 
a monoenergetic beam by an optical element with a straight axis along 
the $z$-direction and comprising a static magnetic field 
$\B = \curl\,\A$.  Hence, the $4$-component spinor wavefunction of 
the beam particle can be assumed to be of the form 
$\Psi(\r,t) = \psi(\r) \exp(-iEt/\hbar)$, where $E$ is the total 
(positive) energy of the particle of mass $m$, charge $q$, and 
anomalous magnetic moment $\mu_a$.  The spatial part of the  
wavefunction $\psi(\r)$ has to obey the time-independent 
Dirac equation 
\be 
H\psi(\r) = E\psi(\r)\,, 
\label{eq:he}
\ee 
where the Hamiltonian $H$, including the Pauli term, is given by 
$$
H = \beta mc^2 + c\val\cdot\hvpi - \mu_a\beta\Vsig\cdot\B\,,
$$
$$
\beta = \left( 
\ba{cc}
\one & \zero \\
\zero & - \one 
\ea \right)\,, \quad  
\val = \left( 
\ba{cc}
\zero & \vsig \\
\vsig & \zero 
\ea \right)\,, \quad 
\Vsig = \left( 
\ba{cc}
\vsig & \zero \\
\zero & \vsig 
\ea \right)\,, 
$$
$$ 
\one = \left(
\ba{cc}
1 & 0 \\
0 & 1 
\ea \right)\,, \ \    
\zero = \left(
\ba{cc}
0 & 0 \\
0 & 0 
\ea \right)\,, 
$$
\be 
\sigma_x = \left( 
\ba {cc}
0 & 1 \\
1 & 0 
\ea \right)\,, \ \  
\sigma_y = \left( 
\ba{cc}
0 & -i \\
i & 0 
\ea \right)\,, \ \ 
\sigma_z = \left( 
\ba{cc}
1 & 0 \\
0 & -1 
\ea \right), 
\label{eq:dh}
\ee 
with $\hvpi = \hp - q\A = -\ih\vnab - q\A$.  Let $p$ be the 
design momentum of the beam particle along the $+z$-direction so that 
$E = +\sqrt{m^2c^4+c^2p^2}$.  The beam is assumed to be paraxial:
$|\pp| \ll |\p| = p$ and $p_z > 0$.

Since we are interested in studying the propagation of the beam 
along the $+z$-direction we would like to rewrite Eq.~(\ref{eq:he}) as 
\be 
\ih\ddz\psi(\rp;z) = \cH\psi(\rp;z)\,.
\label{eq:ch}
\ee
So, we multiply~Eq.~(\ref{eq:he}) from left by $\al_z/c$ and rearrange 
the terms to get 
\bea 
\cH & = & -p\beta\chi\al_z - qA_z\id + \al_z\val_\perp\cdot\hvpip 
+ (\mu_a/c)\beta\al_z\Vsig\cdot\B\,, \nn \\ 
\chi & = & \left( 
\ba{cc}
\xi \one & \zero \\
\zero & - \xi^{-1} \one 
\ea \right)\,, \quad 
\id = \left( 
\ba{cc}
\one & \zero \\
\zero & \one  
\ea \right)\,, \quad 
\xi = \sqrt{\frac{E+mc^2}{E-mc^2}}\,. 
\eea 
Note that the matrix $\beta\chi\al_z$, coefficient of $-p$ in $\cH$, is 
diagonalized as follows: 
\be 
M(\beta\chi\al_z)M^{-1} = \beta\,, \qquad  
M = \frac{1}{\sqrt{2}}(\id + \chi\al_z)\,. 
\ee
Hence, let us define
\be 
\psi' = M\psi\,.
\label{eq:mtr}
\ee
This turns Eq.~(\ref{eq:ch}) into
\be 
\ih\ddz\psi' = \cH'\psi'\,, \qquad 
\cH' =  M\cH M^{-1} = -p\beta + \cE + \cO\,,
\label{eq:ch'}
\ee
with the `even' operator $\cE$ and the `odd' operator $\cO$ given,  
respectively, by the diagonal and off-diagonal parts of 
\be 
\cE + \cO = \left(
\ba{ll}
\ba{l}
\left[-qA_z\one - (\mu_a/2c) \right. \\
\times\left\{\left(\xi + \xi^{-1}\right) 
\vsig_\perp\cdot\B_\perp\right. \\
\left.\left. + \left(\xi - \xi^{-1}\right) 
\sigma_zB_z\right\}\right]
\ea & \ba{l}
\xi\left[\vsig_\perp\cdot\hvpi_\perp - (\mu_a/2c) \right. \\
\times \left\{i\left(\xi - \xi^{-1}\right) \right. \\
\left(B_x\sigma_y - B_y\sigma_x\right) \\ 
\left.\left. - \left(\xi + \xi^{-1}\right)B_z\one\right\}\right] 
\ea \\
\ba{l}
-\xi^{-1}\left[\vsig_\perp\cdot\hvpi_\perp + (\mu_a/2c) \right. \\
\times\left\{i\left(\xi - \xi^{-1}\right) \right. \\
\left(B_x\sigma_y - B_y\sigma_x\right) \\ 
\left.\left. + \left(\xi + \xi^{-1}\right)B_z\one\right\}\right] 
\ea & \ba{l} 
\left[-qA_z\one - (\mu_a/2c) \right. \\
\times\left\{\left(\xi + \xi^{-1}\right)
\vsig_\perp\cdot\B_\perp\right. \\
\left.\left. - \left(\xi - \xi^{-1}\right) 
\sigma_zB_z\right\} \right] 
\ea
\ea \right). 
\ee 

The significance of the transformation in Eq.~(\ref{eq:mtr}) is that 
for a paraxial beam propagating in the forward $(+z)$ direction 
$\psi'$ is such that its lower pair of components are very small 
compared to the upper pair of components, exactly like for a 
positive energy nonrelativistic Dirac spinor.  Note the perfect 
analogy: nonrelativistic case $\rightarrow$ paraxial, positive 
energy $\rightarrow$ forward propagation, and $mc^2 \rightarrow -p$ 
in Eq.~(\ref{eq:dh}) and Eq.~(\ref{eq:ch'}) respectively.  

Let us recall that the FW transformation technique~\cite{FW,BD} is 
the most systematic way of analyzing the standard Dirac equation as 
a sum of the nonrelativistic part and a series of relativistic 
correction terms.  So, the application of an analogous technique 
to Eq.~(\ref{eq:ch'}) should help us analyze it as a sum of the 
paraxial part and a series of nonparaxial correction terms.  To this 
end, we define the FW-like transformation  
\be 
\psi_1 = S_1\psi'\,, \qquad    
S_1 = \exp\left(-\beta\cO/2p\right)\,.  
\label{eq:fwt}
\ee 
The resulting equation for $\psi_1$ is 
\bea 
   &   & \ih\ddz\psi_1 = \cH_1\psi_1\,, \quad 
\cH_1 = S_1\cH'S_1^{-1} - \ih S_1\ddz\left\{S_1^{-1}\right\} 
= -p\beta + \cE_1 + \cO_1\,, \nn \\
   &   & \cE_1 = \cE - \frac{1}{2p}\beta\cO^2 + \cdots\,, \quad 
\cO_1 = -\frac{1}{2p}\beta\left\{\left[\cO,\cE\right] + 
\ih\ddz\cO\right\} + \cdots\,. 
\label{eq:psi1}
\eea 
A series of such transformations successively with the same type of  
recipe as in Eq.~(\ref{eq:fwt}) eliminates the odd parts from $\cH'$ 
up to any desired order in $1/p$.  It should also be mentioned that 
these FW-like transformations preserve the property of $\psi'$ that 
its upper pair of components are large compared to the lower pair 
of components.  We shall stop with the above first step which would 
correspond to the paraxial, or the first order, approximation.  

Since the lower pair of components of $\psi_1$ are almost vanishing 
compared to its upper pair of components and the odd part of $\cH_1$ 
is negligible compared to its even part, up to the first order 
approximation we are considering, we can effectively introduce a  
two-component spinor formalism based on the representation of 
Eq.~(\ref{eq:psi1}).  Naming the two-component spinor 
comprising the upper pair of components of $\psi_1$ as  
$\psi^{\prime\prime}$ and calling the $2 \times 2$ $11$-block element of 
$\cH_1$ as $\cH^{\prime\prime}$ it is clear from Eq.~(\ref{eq:psi1}) 
that we can write  
\bea 
   &   & \phantom{{\rm with}\ \hat{\pi}_\perp^2 = \hat{\pi}_x^2 
+\hat{\pi}} \ih\ddz\psi^{\prime\prime} = 
\cH^{\prime\prime}\psi^{\prime\prime}\,, \nn \\ 
\cH^{\prime\prime} & \approx & \left(-p - qA_z +  
\frac{1}{2p}\hat{\pi}_\perp^2\right) - \frac{1}{p}\left\{(q + \eps) 
B_zS_z + \g\eps\B_\perp\cdot\S_\perp\right\}\,,  \nn \\ 
{\rm with} &   &  
\hat{\pi}_\perp^2 = \hat{\pi}_x^2 + \hat{\pi}_y^2\,, 
\ \  \eps = 2m\mu_a/\hbar\,, \ \ \g = E/mc^2\,, \ \  
\S = \hbar\vsig/2.
\label{eq:hdp}
\eea

Up to now, all observables, field components, etc., are defined 
with reference to the laboratory frame.  But, as is usual in 
accelerator physics, we have to define spin with reference 
to the instantaneous rest frame of the particle while keeping the 
other observables, field components, etc., defined with reference 
to the laboratory frame.  For this, we have to transform 
$\psi^{\prime\prime}$ further to an `accelerator optics representation',  
say, $\psia = T_A\psi^{\prime\prime}$.  The choice of $T_A$ is 
dictated by the following consideration.  Let the operator $\hO$ 
correspond to an observable $O$ in the Dirac representation 
(Eq.~(\ref{eq:dh}) or Eq.~(\ref{eq:ch})).  The operator corresponding 
to $O$ in the representation of Eq.~(\ref{eq:hdp}) can be taken to 
be given by 
\be 
\hO^{\prime\prime} = {\rm the\ hermitian\ part\ of\ the}\ 11{\rm 
-block\ element\ of}\ \left(S_1M\hO M^{-1}S_1^{-1}\right)\,.
\label{eq:hodp}
\ee  
The corresponding operator in the accelerator optics representation 
will be 
\be 
\hOA = {\rm the\ hermitian\ part\ of}\ 
\left(T_A\hO^{\prime\prime}T_A^{-1}\right)\,. 
\label{eq:hoa}
\ee
The operator
\be 
\S^{(R)} = \half\hbar\left(
\ba{cc}
\vsig - \frac{c^2(\vsig\cdot\hvpi\hvpi + 
\hvpi\vsig\cdot\hvpi)}{2E(E+mc^2)} & \frac{c\hvpi}{E} \\
\frac{c\hvpi}{E} & -\vsig + \frac{c^2(\vsig\cdot\hvpi\hvpi + 
\hvpi\vsig\cdot\hvpi)}{2E(E+mc^2)}
\ea \right)\,.
\ee
corresponds to the rest-frame spin in the Dirac representation~\cite{ST}.  
We demand that the components of the rest-frame spin operator in the 
accelerator optics representation be simply the Pauli spin matrices, 
{\em i.e.}, $\S^{(R)}_A \approx \hbar\vsig/2$, up to the first order 
(paraxial) approximation.  This demand leads to the choice 
\be 
\psia = T_A\psi^{\prime\prime}\,, \qquad 
T_A = \exp\left\{-i\left(\hpi_x\sigma_y - 
\hpi_y\sigma_x\right)/2p\right\}\,.
\label{eq:aot}
\ee  
Now, finally, with the transformation given by Eq.~(\ref{eq:aot}), 
the desired basic equation of the quantum beam optics of the Dirac 
particle becomes, up to paraxial approximation, 
\bea 
  &  & \ih\ddz\psia = \cHA\psia\,, \quad  
\cHA \approx \left(-p - qA_z + \frac{1}{2p}\hat{\pi}_\perp^2\right) 
+ \frac{\g m}{p}\Vomeg\cdot\S\,,\nn \\
  &  & \phantom{\ih\ddz\psia = \cHA\psia\,, \quad\quad} {\rm with} \ \  
\Vomeg = -\frac{1}{\g m}\left\{q\B + \eps\left(\B_\parallel + 
\g\B_\perp\right)\right\}\,. 
\label{eq:cha}
\eea 
Here, $\B_\parallel$ and $\B_\perp$ are the components of $\B$ in 
the $+z$-direction (the predominant direction of motion of the beam 
particles) and perpendicular to it, unlike in the usual TBMT vector 
$\Omeg$ in which the components $\B_\parallel$ and $\B_\perp$ are 
defined with respect to the direction of the instantaneous velocity 
of the particle.  The quantum beam optical Hamiltonian $\cHA$ is the 
beam optical version of the DK Hamiltonian in the paraxial approximation.  
To get the higher order corrections, in terms of $\hvpi_\perp/p$ 
and $\hbar$, we have to go beyond the first FW-like transformation.  

It must be noted that while the exact quantum beam optical Dirac 
Hamiltonian $\cH$ (see Eq.~(\ref{eq:ch})) is nonhermitian the nonunitary 
FW-like transformation has projected out a hermitian $\cHA$ 
(see Eq.~(\ref{eq:cha})).  Thus, for the particle that survives the 
transport through the system, without getting scattered far away,  
$\psia$ has unitary evolution along the $z$-axis.  Hence, we can 
normalize the two-component $\psia$, at any $z$, as 
$\left\langle\psia(z)\right.\left|\psia(z)\right\rangle = 
\int\int dxdy\,\psia^\dagger\psia = 1$.  This normalization will be 
conserved along the optic ($z$) axis.  Then, for any observable $O$ 
represented by a hermitian operator $\hOA$, in the accelerator optics 
representation (Eq.~(\ref{eq:cha})), we can define the average at 
the transverse plane at any $z$ as 
\be 
\langle O \rangle(z) = 
\left\langle\psia(z)\right|\hOA\left|\psia(z)\right\rangle  
= \int\int dxdy\,\psia^\dagger\hOA\psia\,. 
\label{eq:av}
\ee
Now, studying the $z$-evolution of $\langle O \rangle(z)$ is 
straightforward.  Integration of Eq.~(\ref{eq:cha}) gives 
\be 
\psia(\zp) = \uzpz\psia(z)\,,
\ee
where the unitary $z$-evolution operator $\uzpz$ can be obtained by 
the standard quantum mechanical methods.  Thus, the relations 
\be 
\langle O \rangle (\zp) = 
\left\langle\psia(z)\right|\uzpzd\hOA\uzpz\left|\psia(z)\right\rangle\,,
\label{eq:otr}
\ee
for the relevant set of observables, give the transfer maps for the 
quantum averages (or their classical values 
{\em \`{a} la} Ehrenfest) from the plane at $z$ to the plane at $\zp$.  
In the classical limit this relation (Eq.~(\ref{eq:otr})) becomes the 
basis for the Lie algebraic approach to classical beam optics~\cite{D}.  
The Lie algebraic approach has been studied~\cite{YEY} in the context 
of spin transfer map also using the classical formalism. 

The main problem of accelerator optics is to know the transfer maps 
for the quantum averages of the components of position, momentum, and 
the rest-frame spin between transverse planes containing the optical 
elements.  We have already seen that the rest-frame spin is 
represented in the accelerator optics representation by the Pauli 
spin matrices.  Let us take that the observed position of the Dirac 
particle corresponds to the mean position operator of the FW 
theory~\cite{FW} or what is same as the Newton-Wigner position 
operator~\cite{T}.  Then, one can show, using Eq.~(\ref{eq:hodp}) 
and Eq.~(\ref{eq:hoa}), that in the accelerator optics representation 
the transverse position operator is given by $\rp$ up to the first 
order approximation (details are given elsewhere~\cite{CJKP2}).  
For the transverse momentum in free space the operator is $\hpp$ in 
the accelerator optics representation.  

\section{An example: the normal magnetic quadru-pole lens} 
For an ideal normal magnetic quadrupole lens of length $L$ 
comprising the field $\B = (Gy,Gx,0)$, corresponding to 
$\A = \left(0,0,\half G\left(y^2 - x^2\right)\right)$, and bounded 
by transverse planes at $z_i$ and $z_f = z_i+L$, the quantum 
accelerator optical Hamiltonian (Eq.~(\ref{eq:cha})) becomes  
\be 
\cHA = \left\{ 
\ba{l}
-p + \frac{1}{2p}\hppsq\,, \quad {\rm for}\ \ z < z_i \ \ {\rm and}\ \ 
z > z_f\,,\\  
-p + \half qG\left(x^2 - y^2\right) + \frac{1}{2p}\hppsq    
- \frac{(q+\g\eps)G\hbar}{2p}\left(y\sigma_x + x\sigma_y\right)\,, \\
\qquad \qquad {\rm for}\ \ z_i \leq z \leq z_f \,.  
\ea \right. 
\ee   
Note that the choice of $\A$ in a different gauge will not affect 
the average values defined by Eq.~(\ref{eq:av}).  Now, using the 
formalism of the previous section, it is straightforward to 
find the desired transfer maps in this case (details are found 
elsewhere~\cite{CJKP1,CJKP2}).  The results are: with 
$\eta = (q+\g\eps)GL\hbar/2p^2$, $K = \sqrt{qG/p}$, $\lambda = h/p$,  
$\langle\ \rangle_i = \langle\ \rangle(z_i)$, and 
$\langle\ \rangle_f = \langle\ \rangle(z_f)$,  
\bea 
   &   & \left( \ba{c}
\xf \\  
\pxf   
\ea \right) \approx \left( 
\ba{cc}
\cos KL & \frac{1}{pK}\sin KL \\
- pK\sin KL & \cos KL 
\ea \right) \nn \\ 
  &   & \quad \qquad \qquad \qquad \times \left(\left( 
\ba{c}
\xin \\  
\pxi  
\ea \right) + \eta\left( 
\ba{c}
\left(\cos KL - 1\right)\siyi/K^2L \\  
- \left(p\sin KL\right)\siyi/KL 
\ea \right)\right)\,, \nn \\ 
   &   & \left( \ba{c}
\yf \\  
\pyf
\ea \right) \approx \left( 
\ba{cc}
\cosh KL & \frac{1}{pK}\sinh KL \\
pK\sinh KL & \cosh KL 
\ea \right) \nn \\ 
  &   & \quad \qquad \qquad \qquad \times \left(\left( 
\ba{c}
\yin \\  
\pyi  
\ea \right) + \eta\left( 
\ba{c}
-\left(\cosh KL - 1\right)\sixi/K^2L \\  
- \left(p\sinh KL\right)\sixi/KL 
\ea \right)\right)\,, \nn \\ 
   &   & \sxf \approx \sxi + \frac{4\pi\eta}{\lambda}\left( 
\left(\frac{\sin KL}{KL}\right) \xszi + \left(\frac{\cos KL - 1}{K^2Lp} 
\right) \pxszi \right)\,, \nn \\ 
   &   & \syf \approx \syi - \frac{4\pi\eta}{\lambda}\left( 
\left(\frac{\sinh KL}{KL}\right) \yszi - \left(\frac{\cosh KL - 1}{K^2Lp} 
\right) \pyszi \right)\,, \nn \\ 
   &   & \szf \approx \szi -  \frac{4\pi\eta}{\lambda}\left\{ 
\left(\frac{\sin KL}{KL}\right) \xsxi - 
\left(\frac{\sinh KL}{KL}\right) \ysyi \right. \nn \\ 
   &   & \qquad \left. + \left(\frac{\cos KL - 1}{K^2Lp}\right) \pxsxi 
+ \left(\frac{\cosh KL - 1}{K^2Lp}\right) \pysyi \right\}.  
\label{eq:map}
\eea 

Obviously we have obtained the well known classical transfer maps 
(matrices) for the transverse phase-space coordinates, and more, 
{\em i.e.}, the transverse SG kicks~\cite{CP,CPP}.  
The longitudinal SG kick, which has been proposed~\cite{CPP} as a 
better alternative to the transverse SG kicks for making a spin-splitter 
device to produce polarized (anti)proton beams, can also be 
understood~\cite{CJKP1,CJKP2} using the present quantum beam optical 
formalism.  The skew magnetic quadrupole can also be analyzed~\cite{K} 
in the same way as here. 

\section{Conclusion}
In summary, we have seen how one can obtain a fully quantum 
mechanical formalism of the accelerator beam optics for a 
spin-$\half$ particle, with anomalous magnetic moment, starting 
{\em ab initio} from the Dirac-Pauli equation.  This formalism leads 
naturally to a unified picture of orbital and spin dynamics 
taking into account the effects of the Lorentz force, the SG   
force and the TBMT equation for spin evolution.  Only the lowest 
order (paraxial) approximation has been considered in some detail,  
with an example.  It is clear from the general theory, presented 
briefly here, that the approach is suitable for handling any magnetic 
optical element with straight axis and computations can be carried 
out to any order of accuracy by easily extending the order of 
approximation.  It should be emphasized that the present formalism 
is valid for all values of design momentum $p$ from the nonrelativistic 
case to the ultrarelativistic case.  The approximation scheme is 
based only on the fact that for a beam, constituted by particles 
moving predominantly in one direction, the transverse kinetic momentum 
is very small compared to the longitudinal kinetic momentum.   

We hope to address elsewhere~\cite{CJKP2} some of the issues related to 
the construction of a more general theory overcoming the limitations of 
the present formalism.  With reference to the inclusion of multiparticle 
dynamics within the present formalism, it might be profitable to be 
guided by the so-called thermal wave model which has been extensively 
developed~\cite{F} in recent years to account for the classical  
collective behavior of a charged-particle beam, by associating with the 
classical beam a quantum-{\em like} wavefunction obeying a 
Schr\"{o}dinger-{\em like} equation with the role of $\hbar$ played by 
the beam emittance $\varepsilon$.  

To conclude, let me emphasize the significance of the quantum 
formalism for beam optics.  The following question has been 
raised: What are the uses of quantum formalisms in beam 
physics?~\cite{C}  Of course, as we have seen above, we understand the 
quantum mechanics underlying the observed classical behavior of 
the beam: when the $\hbar$-dependent quantum corrections are 
worked out through the higher order FW-like transformations it is 
found that they are really negligible.  In my opinion, quantum 
formalism of beam optics has more significant uses.  To see this, 
let me cite the following cases: (1) Recently there is a renewed 
interest~\cite{A} in the form of the force experienced by a spinning 
relativistic particle in an external electromagnetic field.  Such 
studies are particularly important in evaluating the possible 
mechanisms of spin-splitter devices~\cite{CPP}.  A thorough analysis 
based on general Poincar\'{e} covariance, up to first order in spin, 
shows that classical spin-orbit systems can be characterized~\cite{H},  
at best, by five phenomenological parameters.  This just points to 
the fact that spin being essentially a quantum aspect one must 
eventually have a quantum formalism to understand really the high 
energy polarized accelerator beams.  (2)  A look at Eq.~(\ref{eq:av}) 
shows that the form of the transfer map for the quantum averages of 
observables will differ from the form of the corresponding classical 
maps by terms of the type $\langle f(\hO_1,\hO_2,\ldots,)\rangle - 
f(\langle\hO_1\rangle, \langle\hO_2\rangle,\ldots,) \neq 0$; 
the differences are essentially due to the quantum uncertainties 
associated with $\psia$ at the initial $z_i$.  For example, a term 
like $x^3$ in a classical map will become in the corresponding 
quantum map $\langle x^3\rangle = \langle x\rangle^3 
+ 3\langle x\rangle\langle(x - \langle x\rangle)^2\rangle + \langle(x 
- \langle x\rangle)^3\rangle$ which need not vanish even on the axis 
(where $\langle x\rangle = 0$).  It is thus clear that, essentially, 
quantum mechanics modifies the coefficients of the various linear and 
nonlinear terms in the classical map making them dependent on the 
quantum uncertainties associated with the wavefunction of the input 
beam; actually, even terms absent in the classical map will be 
generated in this manner with coefficients dependent on the quantum 
uncertainties as a result of modifications of the other terms (a  
related idea of fuzzy classical mechanics, or fuzzy quantum evolution 
equations, occurs in a different context~\cite{GPC}). This quantum 
effect could be significant in the nonlinear dynamics of accelerator 
beams.  This effect is relevant for spin dynamics too.  For example, it 
is seen in  Eq.~(\ref{eq:map}) that the spin transfer map is not 
linear in its components, in principle, even in the lowest order 
approximation, since terms of the type, say, $\xszi$, $\pxszi$, etc., 
are not the same as $\xin\szi$, $\pxi\szi$, etc., respectively, in 
general.    

\section*{Acknowledgments}
In this First Book on {\it Quantum Aspects of Beam Physics} I would 
like to record my gratitude to Prof. E.C.G. Sudarshan for 
initiating my work on the topic of Dirac spinor beam optics.  It is 
a pleasure to thank Prof. Pisin Chen, and the Organizing Committee 
of QABP98, for sponsoring my participation in the historic Monterey 
meeting and for the hospitality I enjoyed during the conference.  I 
am thankful to our Director, Prof. R. Ramachandran, for kind 
encouragement and my thanks are due to him, and to our Institute, 
for providing full financial support for my travel to participate 
in QABP98.  I wish to thank Prof. Swapan Chattopadhyay for bringing 
to my notice the literature on the application of Lie methods to 
spin dynamics.

\end{document}